\documentclass[aps,prb,twocolumn,superscriptaddress,showpacs]{revtex4}
\usepackage{graphicx}
\usepackage{calc}
\usepackage{bm}

\begin{document}

\title{Experimental realization of a Fabry-Perot-type interferometer by co-propagating edge states in the quantum Hall regime}

\author{E.V.~Deviatov}
\email[Corresponding author. E-mail:~]{dev@issp.ac.ru}
 \affiliation{Institute of Solid State
Physics RAS, Chernogolovka, Moscow District 142432, Russia}

\author{A.~Lorke}
\affiliation{Laboratorium f\"ur Festk\"orperphysik, Universit\"at
Duisburg-Essen, Lotharstr. 1, D-47048 Duisburg, Germany}

\date{\today}

\begin{abstract}
A Fabry-Perot-type interferometer is experimentally realized for electrons in a semiconductor device. A special experimental geometry creates interference conditions for co-propagating electrons in  quantum Hall edge states, which results in oscillations of the current through the device. The  visibility of these oscillations is found to increase at the high-field edge of the quantum Hall plateau. 
\end{abstract}


\maketitle

Interference phenomena are among the most basic manifestations of quantum mechanics, since they directly show the wave properties of the investigated objects. Interference of carriers in semiconductors have therefore attracted considerable interest~\cite{heiblum,goldman,litvin,glattli,feldmanfrac,feldmanabelian}.  In quantum mechanics, the interference takes place if there are several non-distinguishable paths for a particle~\cite{landau}. In semiconductors, the interference scheme was recently realized in Mach-Zehnder type interferometer~\cite{heiblum} by using edge state (ES) transport~\cite{buttiker} in the quantum Hall (QH) effect regime. 

Current-carrying ES are arising at the sample edge at the intersections of the Fermi level and distinct Landau levels~\cite{buttiker}. Split-gates are used to separate ES and bring them into contact in two regions, called as quantum point contacts (QPC)~\cite{heiblum,goldman}. These QPC are resembling semi-transparent mirrors in the Mach-Zehnder optical scheme. The electron's path is divided into two   at the first QPC, and are brought back into contact at the second one. Sweeping of the magnetic field allows to change   phase shift between theses two paths, which gives rise to  interference oscillations of the current through the device~\cite{heiblum,goldman,litvin,glattli}. These oscillations could be used to probe fractional statistics in the fractional QH  regime~\cite{feldmanfrac} or even non-abelian ones for appropriate QH states~\cite{feldmanabelian}.

The major difference between the semiconductor Mach-Zehnder interferometer and the optical device  is the \textit{counter-propagating} interference in the former case. In fact, there is a special direction for the electron propagation in ES, defined by the magnetic field. In QPC, two ES are brought into contact, that are propagating at two opposite gate edges of the QPC (see Fig.~1 in Refs.~\onlinecite{heiblum,litvin}). They therefore form \textit{counter-propagating} paths for the interference in contrast to the optical Mach-Zehnder scheme. A recently reported Fabry-Perot interferometer~\cite{goldmanfabry} is also realized in the counter-propagating scheme.

Because of the chiral nature of ES, there is a principal difference between co- and counter-propagating interference environments, especially in the fractional QH effect regime, where ES are described by the Luttinger liquid picture~\cite{wen}. The interference between co-propagating Luttinger liquids is a new scientific problem both for the theoretical and for the experimental investigations. To realize the interference between \textit{co-propagating} electrons, ES should be used which are moving along the same sample edge, and these ES have to be contacted independently. 

Here, a Fabry-Perot-type interferometer is experimentally realized for electrons in a semiconductor device. A special experimental geometry creates interference conditions for co-propagating electrons in  quantum Hall edge states, which results in oscillations of the current through the device. The  visibility of these oscillations is found to increase at the high-field edge of the quantum Hall plateau. 

\begin{figure}
\includegraphics*[width=0.6\columnwidth]{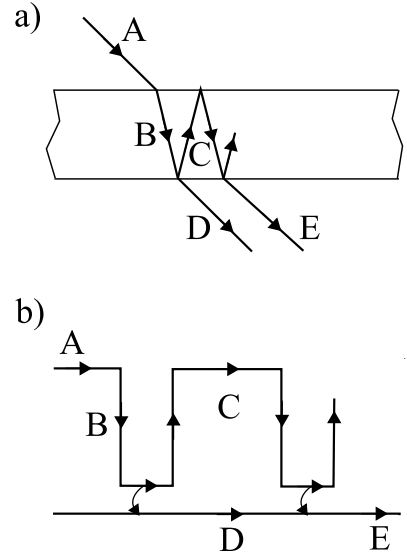}%
\caption{  Schematic diagram of the Fabry-Perot type interferometer, realized as the optical device (a) or the electronic one (b). Only two possible paths are shown for a particle: A-B-D and A-B-C-E. 
\label{fabry}}
\end{figure}

A well-known principle of operation is shown in Fig.~\ref{fabry} (a) for optical Fabry-Perot interferometer. A light beam A, being  refracted into  the media between two mirrors (B), is partially reflected at the other mirror (C) and partially transmitted outside the media (D). The beam C goes in a similar way and is also partially transmitted outside the media (E). If the coherence length for a photon is higher than the geometrical sizes of the media, it is impossible to determine a path for a particular photon (A-B-D or A-B-C-E). The signal \textit{at infinity} is determined by the sum of the probabilities for these paths, and, therefore, depends on the phase shift between them. The latter can be controlled, for example, by the refraction index of the media at constant geometrical sizes of the device.

In Fig.~\ref{fabry} (b) an analogical electron device is shown. Two ES with different electrochemical potentials are brought into  interaction over a small distance $l_{int}$. An electron could be transferred here into the ES with lower potential (the outer one) with low probability, so that the ES are still out of balance after the interaction region. They are then separated for a some distance $w$ and rejoined again. This process repeats itself in the adjacent  interaction region, and so on. For a particular electron, it is impossible to distinguish between the allowed paths (A-B-D-E or A-B-C-E in Fig~\ref{fabry} (b) in analogy with the corresponding optical paths in Fig.~\ref{fabry} (a)), if the electron coherence length exceeds the geometrical distance between the interaction regions $l_c>w$. If the geometrical length  of the interaction region is much smaller than the equilibration length~\cite{mueller}, $l_{int}<<l_{eq}$, multiple electron trajectories are possible. The transport current is determined by the sum of the path's probabilities, and should be sensitive to the phase shift between the paths. The phase shift can be controlled by a magnetic field sweep, that changes the number of flux quanta in the area between the interaction regions and finally produces oscillations of the transport current through the device. This scheme can easily be realized in the quasi-Corbino geometry~\cite{alida}, that provides the independent contacting of ES and measurements of the inter-ES transport current at high imbalances. The original experimental geometry~\cite{alida}  only needs to be modified to have several interaction regions, separated by the small distances.

\begin{figure}
\includegraphics*[width=0.7\columnwidth]{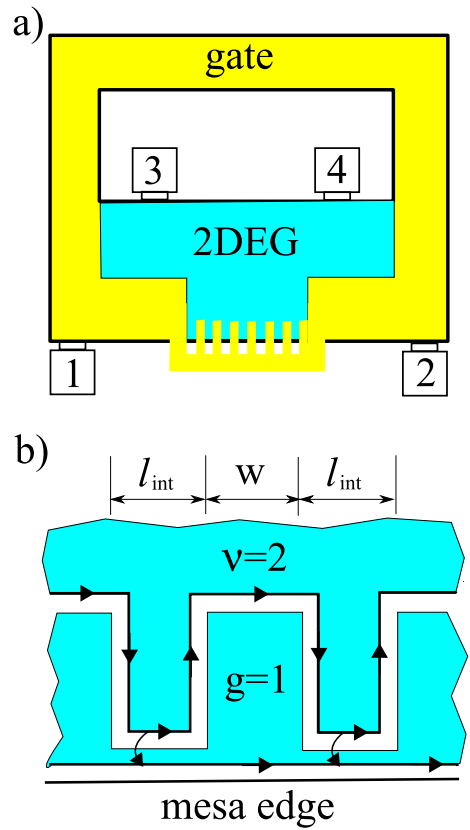}%
\caption{(Color online) (a) Schematic diagram of the sample (not in the scale). The etched mesa edges are shown by thick solid lines. The outer sample dimension is about 1x1~mm$^2$. The inner etched region (white) is approximately 0.5x0.5~mm$^2$. Light yellow (light gray) areas indicate the split-gate, that covers 2DEG around the inner etched region and forms a 10~$\mu$m width gate-gap region at the outer mesa edge. Light green (gray) area indicates uncovered 2DEG. The gate-gap region contains a gate finger structure, connected to the main gate. Ohmic contacts are denoted by bars with numbers. (b) Schematic diagram of a single finger area. The width of each finger is $w=200$~nm, they are separated by the $l_{int}=400$~nm-wide interaction regions. Paths for an electron in the finger region are depicted by the solid lines with arrows. Curved arrows indicate regions of the charge transfer between the ES. Light green (gray) areas are the incompressible regions at filling factors $\nu=2$ (in the bulk) and $g=1$ (under
the gate finger and the incompressible strip at the mesa edge).
Compressible regions (white) are at the electrochemical potentials
of the corresponding ohmic contacts, denoted by bars with numbers in part (a).
\label{sample}}
\end{figure}

Our samples are fabricated from a molecular beam epitaxially-grown GaAs/AlGaAs heterostructure. It contains a 2DEG located 90~nm below the surface. The mobility at 4K is 800 000 cm$^{2}$/Vs  and the carrier density 4.4 $\cdot 10^{11}  $cm$^{-2}$, as it was obtained from standard magnetoresistance measurements. 

Samples are patterned in the quasi-Corbino sample
geometry~\cite{alida} with additional gate fingers structure in the gate-gap region, see Fig.~\ref{sample}, (a). 
The sample has an etched region
inside, providing a topologically independent inner and outer  mesa edges (Corbino topology). In a quantizing magnetic field, at filling factor $\nu=2$, two ES are arising near each mesa edge. These ES are at the electrochemical potential of the corresponding ohmic contacts~\cite{buttiker}. A split-gate is used to redirect one of the ES to the other mesa edge, see Fig~\ref{sample} (b), by partially depleting the 2DEG to filling factor $g=1$.  The gate-gap region at the outer mesa edge is of microscopic size  (10~$\mu$m in the present samples), that is smaller than the equilibration length (about 1~mm for spin-split edge states~\cite{mueller} at low temperature), in contrast to the macroscopic ungated region at the inner mesa edge. For these reasons, applying a
voltage between ohmic contacts at outer and inner edges leads to the electrochemical potential imbalance between the two ES in the gate-gap region, equal to the applied voltage~\cite{alida}. 

A special structure of the gate fingers is patterned in the gate-gap region, to obtain the Fabry-Perot geometry (\textit{cf.} Figs.~\ref{fabry} (b),\ref{sample}). Each finger ($w=200$~nm wide) is at the gate potential. It depletes the 2DEG beneath to filling factor $g=1$ and therefore decouples the two ES in the finger region. Inter-ES transport is allowed in the regions between the fingers only, see Fig~\ref{fabry} (b). Each region is of width $l_{int}=400$~nm, which is much smaller than the equilibration length~\cite{mueller} $l_{eq}\sim 1$~mm. This ensures  low probability of electron transfer in a particular region, and, therefore, several possible paths for an electron. The whole structure contains 14 fingers in the gate-gap. The finger length was lithographically defined to overlap with the mesa edge. The exact effective length should be determined from the experiment, because of the finite depletion region at the etched mesa edge. Some samples have no finger structure in the gate-gap and are used for reference. 

We study $I-V$ curves in 4-point configuration, by applying a \textit{dc} current between one pair of inner and outer contacts ($\sim$100~Ohms) and measuring the \textit{dc} voltage between another pair of inner and
outer contacts. The contact behavior is tested separately by  2-point magnetoresistance measurements.  The measurements are performed in a dilution refrigerator with a base temperature of 30~mK, equipped with a superconducting solenoid. The results, presented here, are independent of the cooling cycle. 

\begin{figure}
\includegraphics[width=0.9\columnwidth]{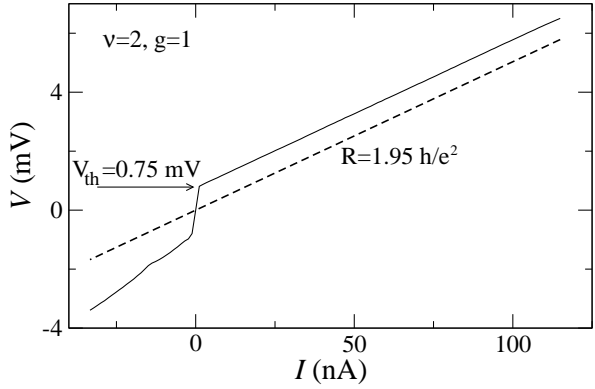}%
\caption{ $I-V$ curve (solid), measured at the center of the $\nu=2$ QH plateau ($B=9.19$~T). The positive branch is linear with a slope $R=1.95 h/e^2$, as demonstrated by the dashed line. The non-linear negative branch is shown around the zero region only.  \label{IV}}
\end{figure}

The experimental $I-V$ curve is shown in  Fig.~\ref{IV} for the filling factors $\nu=2, g=1$.  It exhibits the common $I-V$ characteristic~\cite{alida}  for this experimental geometry: the positive branch starts from the threshold voltage $V=V_{th}$ and exhibits a constant slope $R$ for $V>V_{th}$, while  the negative branch is strongly non-linear.  The  slope of the positive branch ($R=1.95 h/e^2$) is close to the equilibrium slope ($R_{eq}=2 h/e^2$ for this contact configuration). $V_{th}$ is defined by the flat-band situation at the edge and corresponds to the exchange-enhanced Zeeman splitting (see Refs.~\onlinecite{alida} for details). 

\begin{figure}
\includegraphics[width=\columnwidth]{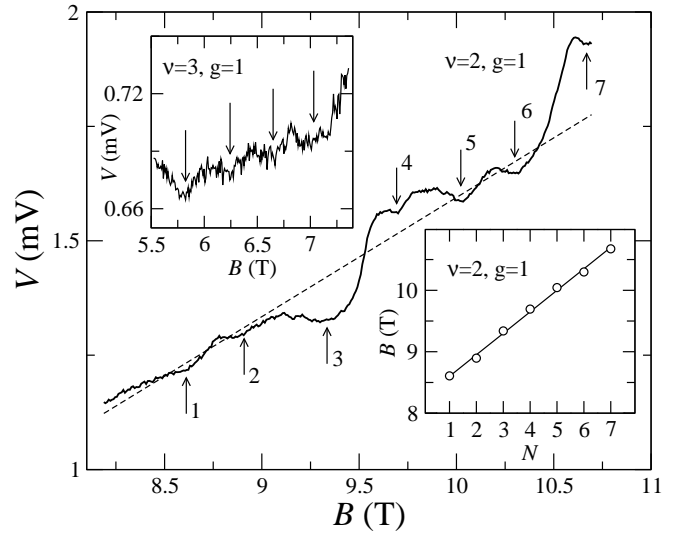}%
\caption{ Main figure: oscillating behavior of the measured voltage while sweeping the magnetic field at $\nu=2,g=1$. Dashed line  indicates the monotonic behavior of the exchange-enhanced Zeeman splitting~\protect~\cite{vadik}. Arrows with numbers show positions of the minima of the oscillations. The measurement current is $I_0=11.49$~nA. The range of the $\nu=2$ QH plateau is 8.2-10.2~T for the present sample. Right inset:  positions of the oscillations versus their index (open circles). The solid line fits the experimental points with the oscillation period $\Delta B=0.35$~T.  Left inset: oscillations at the filling factors $\nu=3,g=1$. The range of the $\nu=3$ QH plateau is 5.8-6.5~T.  \label{oscill}}
\end{figure}

To observe the interference effects, we fix the measurement current to $I_0=11.49$~nA (i.e. at some point above the threshold) and sweep the magnetic field around the $\nu=2$ QH plateau.  The result is depicted in Fig.~\ref{oscill}: clear visible oscillations of the measured voltage are present, instead of monotonic  increase (caused by the linear increase of the spin gap, see dashed line), that would be present without the finger structure. These oscillations are better pronounced at the high-field edge of the QH plateau, and even beyond it, while the visibility is diminishing towads lower fields. This oscillating behavior does not depend on the measurement current: it has been tested that the  positions and the amplitudes of the oscillations are the same at any current above the threshold. This means that it is the threshold voltage that is sensitive to the magnetic field sweep and not the slope of the positive branch. This fact is confirmed in direct measurements of the $I-V$ curves at different magnetic fields in the range of Fig.~\ref{oscill}.

The right inset to Fig.~\ref{oscill} shows the  positions of the minima as a function of their index. The experimental data follow a straight line, that gives the period of the oscillations $\Delta B=0.35$~T.  This period corresponds to the $2\pi$ phase shift between electron paths around one gate finger, see Fig.~\ref{fabry}. In other words, it corresponds to the change of the  magnetic flux through the finger region  by one flux quantum $\Phi_0$, that allows to estimate the effective finger area to $1.2\cdot10^{-2} \mu\mbox{m}^2$. 

The same behavior can be seen for filling factors $\nu=3, g=1$, see left inset in Fig.~\ref{oscill}, that corresponds to the same spin-split ES, as in the $\nu=2, g=1$ case. Oscillations are clearly visible, despite higher noise at low signals. In contrast, no oscillations are found for the filling factors $\nu=3, g=2$, that corresponds to the transport between the cyclotron-split ES, even at much higher signals~\cite{alida}. We also tested that there are no oscillations at any filling factors for the reference samples without gate finger structure in the gate-gap.

Let us start the discussion from the $\nu=2, g=1$ situation. The transport takes place between two spin-split ES in the regions between the fingers. The slope of the positive $I-V$ branch depends only on the electrochemical potential redistribution~\cite{alida}, and thus is not sensitive to the phase shift around the finger region. Instead, it is the threshold voltage that is varied in dependence on the phase shift: the interference can facilitate or inhibit the starting condition for the current flow. It means, that the magnetic field sweep produces vertical shifts of the positive branch above the threshold, with constant equilibrium slope of the linear part. This is exactly what we  see in the experiment. It explains the independence of the oscillations of the measurement current and the particular contact configuration (the latter affects only on the slope of the $I-V$ curve~\cite{alida}). It also explains why the oscillations are situated around the monotonic exchange-enhanced Zeeman gap~\cite{vadik}. 

The interference oscillations in transport current indicate, that the electron's coherence length $l_c$ exceeds~\cite{coherence} the finger perimeter $\sim w \sim 200$~nm.  The coherence length is a new and independent parameter of the problem. It reveals itself in the visibility of  the oscillations  and thus should be discussed in detail.

The real sample edge potential is smooth, and this gives rise to the compressible-incompressible strips formation~\cite{shklovsky}, see Fig.~\ref{sample} (b). Landau levels are pinned to the Fermi level in some regions (compressible strips), while the filling factor is constant in others (incompressible ones). The dissipativeless (diamagnetic) current, flowing along the sample edge, is carried in the incompressible regions only, because the group velocity is zero in compressible regions. Out of balance, however, the border position between the compressible and incompressible strip is changed. It is the electrons in this region that define the transport current~\cite{tauless,gerhards}, that supports our simple ES structure in Fig.~\ref{sample} (b). 

Because of backscattering suppression in the QH state~\cite{buttiker}, transport along the sample edge has a little influence on the phase coherence. The transport in the direction, perpendicular to the sample edge is carried by tunneling across the incompressible strip and by diffusion within the compressible strip. The presence of the compressible regions has a crucial influence on the phase coherence, restricting the coherence length.  It allows to explain the growth of the oscillations visibility while moving beyond the high field edge of the QH plateau. Above the plateau, there is the same incompressible strip at the sample edge, as in Fig~\ref{sample} (b), while the bulk of the sample is in the compressible state. The compressible bulk state screens the electric fields, which narrows the edge regions and, therefore, the compressible regions widths~\cite{gerhards}. Thus, the coherence length is a maximum at the high-field edge of the QH plateau. The coherence length dependence on  the compressible regions, could also destroy some of the proposed experiments on non-abelian statistics~\cite{feldmanabelian}.

The compressible-incompressible strips formation also defines the effective gate finger area. It  no longer coincides with the lithographic dimensions. Namely, the width of the finger at any side should be diminished on the approximately half-distance between the gate and the 2DEG. That gives the effective width $w_{eff}$ equals to $\approx 0.1 \mu$m. because of this small $w$ value, the effective finger length is of the same order, irrespective of the lithographic overlap with the mesa edge. The effective area is therefore about $\sim w_{eff}^2$, that is in good correspondence with our experimental observation. It is clear now, that they are the geometrical dimensions that restrict the visibility in the present device. To improve the visibility, $w$ should be seriously increased, while it's better to diminish $l_{int}$.

The situation at the filling factors $\nu=3, g=1$ is similar to the one discussed above. At this filling factor combination transport takes place between the same spin-split ES, as in the $\nu=2, g=1$ case. It is not the case at $\nu=3, g=2$: the transport in this case takes place between spin-split sublevels of the \textit{different} Landau levels. The paths in Fig.~\ref{fabry} (b) becomes to be non-equivalent in this case,  which could destroy the interference.

We wish to thank  V.T.~Dolgopolov for fruitful discussions and A.~W\"urtz for sample preparation.
We gratefully
acknowledge financial support by the RFBR, RAS, the Programme "The
State Support of Leading Scientific Schools". E.V.D. is also supported by MK-1678.2008.2.


\begin{thebibliography}{99}
\bibitem{coherence} $l_c$ is about the sample dimention $\sim 10 \mu$m in Refs.~\protect\onlinecite{heiblum,goldman,litvin,glattli} that supports our estimation.
\end{thebibliography}
\end{document}